\documentclass[10pt,journal,compsoc]{IEEEtran}

\ifCLASSOPTIONcompsoc
  \usepackage[nocompress]{cite}
\else
  \usepackage{cite}
\fi

\usepackage{graphicx}
\usepackage{array}
\usepackage{makecell}
\usepackage[dvipsnames]{xcolor}
\usepackage{amsfonts}
\usepackage{amssymb} 

\usepackage{url}

\usepackage{xcolor}
\usepackage{todonotes}
\usepackage{comment}
\usepackage{subfig}

\begin{document}
	
\newcolumntype{L}[1]{>{\raggedright\arraybackslash}p{#1}}
\newcolumntype{C}[1]{>{\centering\arraybackslash}p{#1}}
\newcolumntype{R}[1]{>{\raggedleft\arraybackslash}p{#1}}

\title{The Pixels and Sounds of Emotion:\\ General-Purpose Representations\\ of Arousal in Games}

\author{Konstantinos Makantasis,
        Antonios Liapis~\IEEEmembership{Member,~IEEE}, and\\ Georgios N. Yannakakis,~\IEEEmembership{Senior~Member,~IEEE}
\IEEEcompsocitemizethanks{\IEEEcompsocthanksitem K. Makantasis, A. Liapis and G. N. Yannakakis are with the Institute of Digital Games, University of Malta, Msida 2080, Malta (e-mail: konstantinos.makantasis@um.edu.mt, antonios.liapis@um.edu.mt, geor- gios.yannakakis@um.edu.mt).\protect\\}
\thanks{Manuscript received Month Day, Year.}}


\IEEEtitleabstractindextext{
\begin{abstract}
What if emotion could be captured in a general and subject-agnostic fashion? Is it possible, for instance, to design general-purpose representations that detect affect solely from the pixels and audio of a human-computer interaction video? 
In this paper we address the above questions by evaluating the capacity of deep learned representations to predict affect by relying only on audiovisual information of videos. We assume that the pixels and audio of an interactive session embed the necessary information required to detect affect. We test our hypothesis in the domain of digital games and evaluate the degree to which deep classifiers and deep preference learning algorithms can learn to predict the arousal of players based only on the video footage of their gameplay. Our results from four dissimilar games suggest that general-purpose representations can be built across games as the arousal models obtain average accuracies as high as $85\%$ using the challenging leave-one-video-out cross-validation scheme. The dissimilar audiovisual characteristics of the tested games showcase the strengths and limitations of the proposed method.
\end{abstract}

}

\maketitle

\IEEEdisplaynontitleabstractindextext

%
\IEEEpeerreviewmaketitle

\IEEEraisesectionheading{\section{Introduction}\label{sec:introduction}}

\IEEEPARstart{D}{esigning} autonomous agents capable of performing equally well across different tasks is a long term vision of artificial intelligence (AI) \cite{goertzel2007artificial}. Towards realizing such a vision, the recent groundbreaking study of Minh \emph{et al.} \cite{mnih2015human} introduces the idea of \emph{general-purpose} deep-learned representations for controlling agents capable of performing well across different tasks. These agents, in particular, managed to achieve superhuman performance in playing 2D Atari games by merely observing the pixels of the screen. As impressive as such a result might be, the derived agents are restricted to act in a particular set of deterministic environments and achieve a clearly- and objectively-defined goal: to maximize their score. Arguably, however, several of the most interesting problems that AI is requested to solve---such as emotion recognition and artificial psychology---have ill-posed and subjectively-defined target functions under non-deterministic contexts.

Inspired by the core principles of Mnih \emph{et al.} \cite{mnih2015human}, in this paper we transfer and introduce the idea of general-purpose representations to the field of affective computing. We thus reframe the user-specific way in which affective detection normally operates and, instead, we investigate the degree to which general-purpose representations can learn to predict emotion. As videos of interaction capture a user's behavior, we base our investigations on the assumption that the audiovisual information contained in such a video can hold information about both the interaction context and the elicited affective patterns, and thus it can be a predictor of the user's experience. Our key hypothesis is that we can construct accurate models of affect based only on the audiovisual information of videos of interaction; as in \cite{mnih2015human}, we test this hypothesis in the domain of games. In particular we attempt to predict a game's arousal level relying solely on the audiovisual information of game footage. 

\begin{figure}[!tb]
	\begin{minipage}{0.5\linewidth}
		\centering
		\centerline{\includegraphics[width=0.98\linewidth]{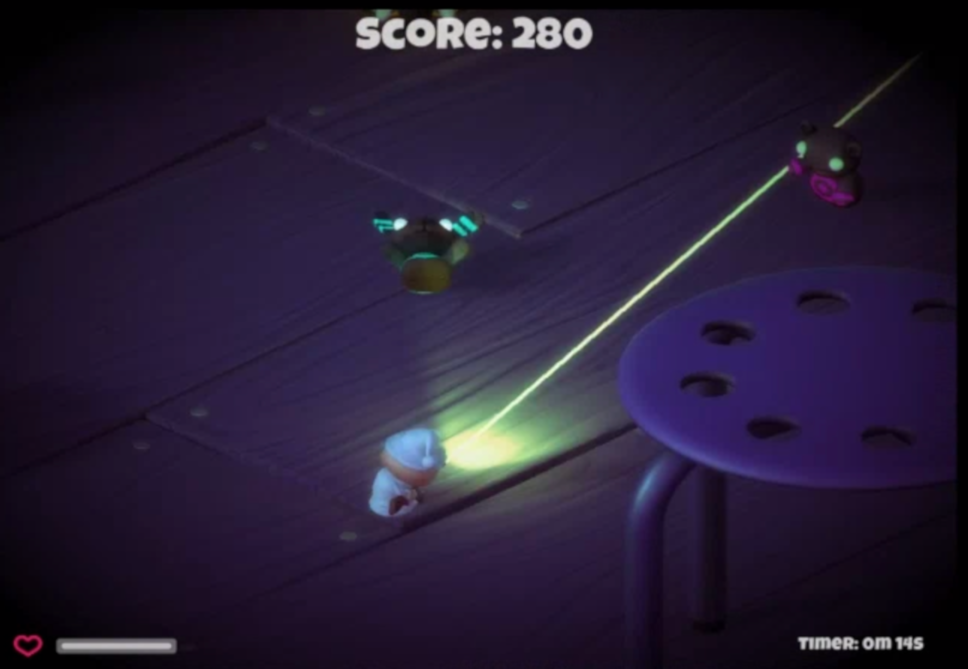}\vspace{0.1cm}}
	\end{minipage}
	\begin{minipage}{0.5\linewidth}
		\centering
		\centerline{\includegraphics[width=0.98\linewidth]{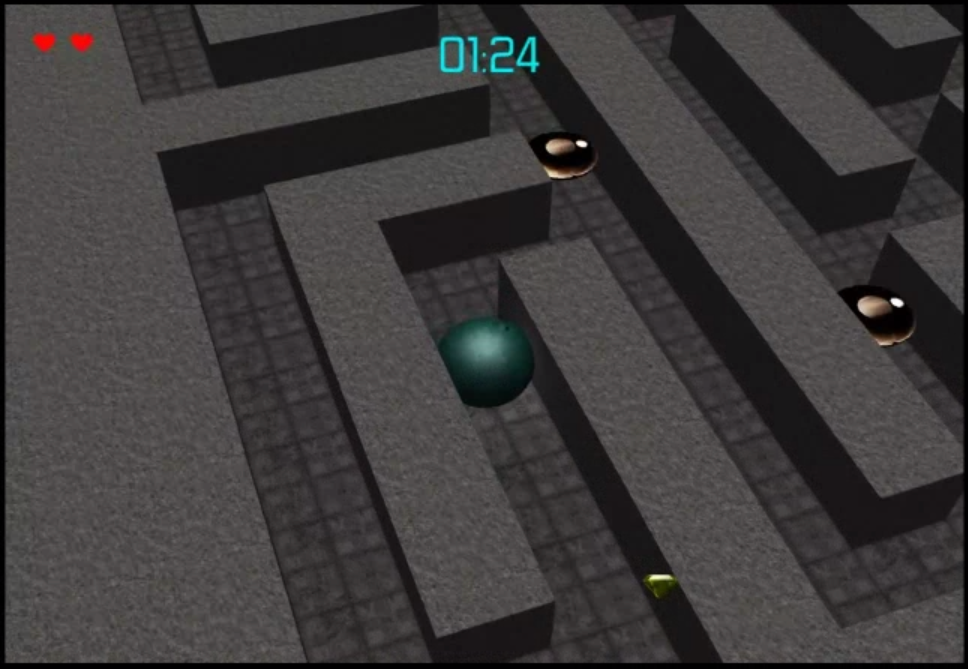}\vspace{0.1cm}}
	\end{minipage}
	\begin{minipage}{0.5\linewidth}
		\centering
		\centerline{\includegraphics[width=0.98\linewidth]{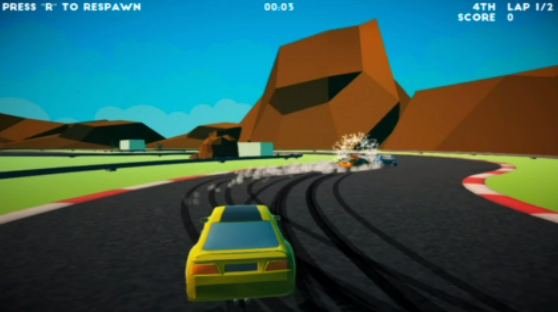}}
	\end{minipage}
	\begin{minipage}{0.5\linewidth}
		\centering
		\centerline{\includegraphics[width=0.98\linewidth]{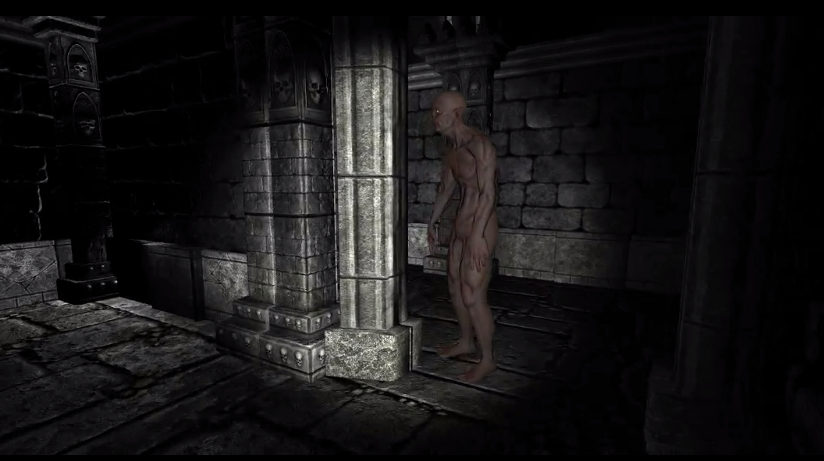}}
	\end{minipage}
	\caption{Screenshots from Survival Shooter (top left), Maze-Ball (top right), Solid Rally (bottom left) and Sonancia (bottom right) games used in this study.}
	\label{fig:game_screenshots}
\end{figure}

Games provide complex yet well-defined environments, which are designed to elicit increased levels of player engagement and motivation \cite{melhart2019your,yannakakis2018artificial}. Players, during their interaction with games, produce gameplay footage that has the unique property of overlaying the game context onto aspects of playing behavior and affect; this suggests that players' affect is embedded in the gameplay context. That embedding, in turn, renders the explicit fusion of context with affect manifestation unnecessary---a dominant practice within affective computing \cite{mcduff2014internet, ringer2018deep, zeng2009survey, zafeiriou2017deep}. 
Although we focus on the games domain, our approach is general and potentially applicable to different human-computer interaction domains, since it relies on raw audiovisual information. Such information fuses the interaction context with the affect of the user as manifested during the interaction.

Given the bimodal (audio and visuals) nature of the affect modeling task, we use a two-stream deep neural network---both as a classifier and as a preference learner---that considers audiovisual gameplay footage information and predicts the player's annotated arousal. The first stream is a Convolutional Neural Network (CNN) that processes visual information as pixels of video frame sequences. The second stream is a fully connected network that processes the game audio information of the considered sequence of video frames. Via late fusion, we propagate the audiovisual information to a fully connected network that predicts arousal. We test the methodology across four dissimilar 3D games and their corresponding gameplay footage (see Fig. \ref{fig:game_screenshots}). All gameplay videos have been annotated by the players themselves (first-person) in terms of arousal using the \textit{RankTrace} \cite{lopes2017ranktrace} continuous annotation tool. Our experimental evaluation validates our hypothesis in most games and suggests that we can derive highly accurate models of affect using general-purpose representations that rely solely on audiovisual information of the interaction. In particular, the two representations (deep classifier and preference learner) predict arousal for three of the games examined with an average classification accuracy that reaches between $82\%$ and $83\%$ on average using the demanding leave-one-video-out cross-validation scheme. The under-performing models in one of the games lead to an insightful discussion regarding the limitations of the proposed approach and the environments it is best suited for.

Our work is novel in several ways. First, we derive accurate models of affect in different games without relying on any direct manifestation of emotion or modality of user input. Instead, our work is one of the first approaches towards modeling players' affect through general-purpose representations of information embedded solely in the context of interaction. Our methodology, thereby, yields affect models that are general and user-agnostic. Second, to the best of our knowledge, this study is the first attempt to derive a function that maps directly audiovisual gameplay information---such as pixels and audio features---to game experience across different games. Finally, via the employed two-stream deep network, we investigate the degree to which each modality, as well as their fusion, can be used as a predictor for such a mapping in affective computing. The paper builds upon and significantly extends the preliminary results of Makantasis \emph{et al.} \cite{makantasis2019pixels}, which map the visual information of gameplay footage to players' arousal in one game. Specifically, in this paper we explore two different modalities of the game footage: besides the visual information we also exploit audio information in an attempt to yield more accurate representations of arousal. Moreover, we approach the arousal modeling problem using two different learning paradigms---a binary classification and a preference learning task---and we compare their performance quantitatively and their top-performing arousal models qualitatively. 
Finally, we test the generality of the proposed methodology across four heterogeneous games with regards to both the audiovisual information they offer to the deep learner and the arousal patterns they elicit.

The remainder of the paper is organized as follows. Section \ref{sec:related_work} reviews related work regarding affect modeling in games and videos. Section \ref{sec:dataset} describes the games, the employed datasets and the data pre-processing approach we followed. Section \ref{sec:methodology} lists the key elements of our methodology including the architectures of the learning models for both binary classification and preference learning. In Section \ref{sec:experiments} we experimentally validate and analyse our models across the four games. Finally, Section \ref{sec:discussion} discusses our main findings and Section \ref{sec:conclusions} concludes this study.

\section{Related Work}
\label{sec:related_work}

This section surveys key literature on affect modelling relevant to the proposed approach of mapping audiovisual data from gameplay videos for predicting affect.

\subsection{Models of Affect Based on Audiovisual Information}\label{sec:audiovisual}

Audiovisual information has been at the core of interest for both eliciting and modeling emotions in affective computing \cite{picard1995affective, bryant2008vocal,calvo2015oxford}. Typically, videos feature the face or the body of one or more humans and their emotions are modeled via non-verbal (visual and vocal) cues \cite{pini2017modeling,calvo2010affect} due to theoretical frameworks and evidence supporting that such modalities can convey emotion \cite{pini2017modeling, ambadar2005enigmatic, bassili1979emotion, ekman1992argument,calvo2010affect}. Visual information is related to the dynamic patterns of human face(s) modeled via facial cues \cite{littlewort2011cert, bartlett2006facial}, body postures \cite{kleinsmith2007posture, kleinsmith2011posturedb}, gestures \cite{glowinski2008gesture} or gait \cite{montepare1987gait, li2016emotion}. Vocal information relies on audio signals which are used to construct acoustic and voice quality cues based on the pitch, the energy, the frequency and the spectrum of the signal \cite{scherer2003vocal, han2006efficient}.

A number of earlier studies base the construction of affect models on ad-hoc features of an image. Indicatively, Liu \emph{et al.} \cite{liu2014combining} combined traditional hand-crafted image features such as SIFT \cite{lowe2004distinctive} and Histogram of Oriented Gradients \cite{dalal2005histograms} as inputs to machine learning models for emotion recognition in the wild. Yao \emph{et al.} \cite{yao2015capturing} hand-crafted image features based on Local Binary Patterns \cite{ahonen2004face} for facial image emotion recognition. Recent advances in deep learning, however, enable the automatic construction of features via convolutional neural networks (CNNs); CNNs were first applied in \cite{baveye2015videoprediction} to predict dimensional affective scores from videos, but the small scale datasets challenged the training of deep models of affect. The need for effectively training CNNs triggered the development of medium- and large-scale affect datasets such as the Celebrity Face in the Wild \cite{zhang2012finding}, the Facial Expression Recognition 2013 Dataset \cite{goodfellow2013challenges} and the Aff-Wild database \cite{zafeiriou2017aff}. Based on these datasets, Ng \emph{et al.} \cite{ng2015deep} used transfer learning and CNNs for emotion recognition through visual cues and Kollias \emph{et al.} \cite{kollias2017recognition} combined CNNs with recurrent neural networks to model arousal and valence. Finally, in \cite{mcduff2014internet} facial expressions were fused with videos of advertisements for predicting whether viewers liked the videos or not.

Regarding emotion recognition via audio data, Eyben \emph{et al.} \cite{eyben2015geneva} conducted a detailed study on audio emotion features. The authors constructed GeMAPS, a concise feature set with 62 audio features. Recent studies show that fusing audio and visual information results in more accurate models than those of a single modality \cite{ding2016audio}. In \cite{ebrahimi2015recurrent} energy and spectral audio features are fused with visual information for emotion prediction in short video clips, while in \cite{tzirakis2019real} audiovisual data is used to train a deep neural network for recognizing affect in real-world environments.

The approach presented in this paper can be seen as unconventional for modeling affect. Following our preliminary study \cite{makantasis2019pixels} on general-purpose pixel-based models of affect, in the current study we use audiovisual information of human-computer interaction as the sole input for modeling the affect of the human across different tasks (i.e. games). The role of the audiovisual interaction footage is thus twofold: the audiovisual information contained in the footage is used to model affect as the context that both elicits and manifests emotion without the need of other external manifestations of affect. The proposed approach is a general method for modeling affect solely via videos containing sound and do not contain either facial/bodily expressions or vocal cues of humans. The experimental validation of the proposed approach---at least within the games domain---suggests that this \emph{subject-agnostic} perspective is not only possible, but it also yields highly accurate models of affect in games with particular audiovisual characteristics.  

\subsection{Affect Modeling in Games}\label{sec:affectivegames}

Affect modeling within the domain of games refers to modeling the behavior and the affective responses of players \cite{yannakakis2012playermodeling,yannakakis2018artificial}. A player model receives as input a modality (or a set of modalities) regarding the player, such as gameplay data and/or physiological measurements, and attempts to predict aspects of the in-game behavior or the player experience. Indicatively Pedersen \emph{et al.} \cite{pedersen2009experience} combined gameplay data (e.g., number of deaths) with game level features to predict players' reported affect using \textit{Super Mario Bros} (Nintendo 1985) as a testbed. Shaker \emph{et al.}  \cite{shaker2013visual} used the same testbed to predict players' affect based on players' posture during gameplay. Recently, Melhart \emph{et al.} \cite{melhart2020moment} managed to successfully model the moment-to-moment engagement level of \emph{PUBG} (PUBG Corporation, 2017) streamed videos by considering the chatting behavior of its viewers. Martinez \emph{et al.} used various deep learning methodologies to capture player experience via gameplay metrics and physiology \cite{martinez2013learning,martinez2014don}. Finally, Camilleri \emph{et al.} \cite{camilleri2017generalmodels} attempted to create arousal models that are general across different games relying solely on gameplay metrics. 

This study advances the state of the art in player modeling as the proposed model of affect is based solely on the audiovisual information contained in gameplay footage. The majority of studies that analyze and extract information from gameplay footage rely on contextual information about the game such as structural and game level elements, physics and mechanics of the game (e.g. \cite{shaker2013visual,guzdial2016deep}). Moreover, the most common approaches for analyzing player experience heavily rely on direct measurements from players under well-defined experimental settings; the modalities that are usually considered include facial expression and head pose \cite{shaker2013visual}, speech \cite{kannetis2009fantasy} and physiology \cite{martinez2013learning,yannakakis2016psychophysiology}.

Building upon and significantly extending the preliminary study of Makantasis \emph{et al.} \cite{makantasis2019pixels}, our methodology models players' experience without any \textit{a priori} contextual knowledge about the game. Instead, it uses general-purpose deep learned representations of gameplay footage (i.e. pixels and audio files) as it ignores functional aspects of the game per se. As a result, our approach does not require any direct in-game feature or manifestation of affect (e.g. via physiology, speech, or facial expression), it is not intrusive, and it allows the rapid collection and processing of vast amounts of data. As gameplay videos are largely available online in massive quantities---e.g. via service such as Twitch\footnote{\url{https://www.twitch.tv/}} and Mixer\footnote{\url{https://mixer.com/microsoft}}---the proposed approach is potentially generalizable to any game with available audiovisual content. 

\subsection{Video Affective Content Analysis}

Video affective content analysis has been an active research area focusing on classifying and retrieving videos based on their affective content. While conventional content-based video analysis relies on generic semantic content, video affective content analysis tries to identify videos that elicit certain emotions in their viewers \cite{wang2015video}. Recent research adopts either direct or implicit approaches. Direct approaches infer the affective content of videos directly from the related audiovisual features, while implicit approaches detect affective content from videos based on an automatic analysis of a user's spontaneous response while consuming the video \cite{baveye2017affective}. Below we discuss direct approaches since they are closely related to the present study.

Hanjalic and Xu \cite{hanjalic2005affective} proposed one of the first direct approaches for video affective content analysis, using handcrafted features of audio and visual information of video segments to model arousal and valence. Since then, extracting audiovisual features and exploiting machine learning methods to model emotion is the most common practice in video affective content analysis \cite{canini2012affective,cui2013mutual,yazdani2013multimedia,xu2013hierarchical}. More recent work takes advantage of deep learning to automatically generate deep features to describe audiovisual information, such as features of motion and scene cues \cite{yi2019multi}. Wang \emph{et al.} \cite{wang2019capturing} use a generative adversarial network to classify emotion of videos, while Mitenkova \textit{et al.} \cite{mitenkova2019valence} use the output of a pretrained network on face images \cite{parkhi2015deep} as input to a tensor regression layer for prediction arousal and valence levels. Zhu \emph{et al.} \cite{zhu2020affective} propose a multimodal deep quality embedding network and a deep learning affective classifier to efficiently process noisy affect data.

Although our study relates to video affective content analysis studies, it is conceptually different. Video affective content analysis tries to model and predict the emotion elicited by a video to a viewer. In contrast, this paper focuses not on the content creator's side, as we aim to model the emotional state of a player while they are playing the game.

\section{Dataset and Data Processing} \label{sec:dataset}

To test the performance and the generality of the proposed approach we used frames and sound from four dissimilar games: \emph{Survival Shooter}, \emph{Maze Ball}, \emph{Solid Rally} and \emph{Sonancia}. Figure \ref{fig:game_screenshots} shows a screenshot of each game. In this section, we describe the games, the datasets obtained from these games and the data cleaning process we followed. Participants were recruited via snowball sampling and were primarily university students who are casual gamers and/or follow courses in game design and ICT, with no prior experience in affect annotation. A different set of participants was used for \emph{Solid Rally} and \emph{Sonancia} whereas the same set of participants was used for \emph{Survival Shooter} and \emph{Maze Ball} \cite{camilleri2017generalmodels}. Prior to annotation, all participants were presented with an introductory screen that describes arousal as ``the intensity of gameplay no matter whether you like the game or not. High arousal can be a feeling of readiness, tension, excitement or exhilaration. Low arousal can be a feeling of fatigue, boredom, calmness or relaxation''. 

\subsection{Testbed Games}

To test how general-purpose input representations can be used for modelling affect, we selected the four games due to their dissimilarities. The games belong to different genres, with different mechanics, camera perspective, pace, visual and audio design. Specifically, Survival Shooter is a fast-paced shooter game that requires accurate aiming and constant movement. Maze Ball is a slow-paced physics game that needs accurate timing of movement. Sonancia is a horror game that elicits negative emotions and disorientation. Finally, Solid Rally is a fast-paced racing game that simulates a multi-player experience with AI drivers. Moreover, the camera perspective is top-down in Survival Shooter, third-person in Maze Ball and first-person in Sonancia and Solid Rally. The dissimilarities between the four games make them ideal for testing the degree to which accurate models of affect can be based on general-purpose input representations. We should also highlight that different sets of players played and annotated three of the four games.

\subsubsection{Survival Shooter}

Survival Shooter (SS) \cite{camilleri2017generalmodels} tasks the player to shoot down as many hostile toys as possible while avoiding collisions with them. Hostile toys spawn at predetermined areas of the level and move towards the player's avatar. The avatar is equipped with a laser gun, which can kill a toy with a few shots. Every toy killed adds to the player's score. Background music plays throughout the gameplay of SS; while the player is firing the laser gun, the volume of music lowers, and the dominant sound is the weapon sound. Sound effects play when the avatar collides with hostile toys, when a toy is killed, and when the player runs out of life. The duration of the gameplay is 60 seconds.

The SS data used in this study was collected from 25 different players (10 females) aged from 19 to 54 (median age 24). Most players considered themselves good or expert players (70\%) while the rest considered themselves novice or non-gamers. Each player played the game and then annotated her recorded gameplay footage in terms of arousal; this play-annotation cycle occurred twice. The first-person annotation process was carried out using the \emph{RankTrace} tool \cite{lopes2017ranktrace,melhart2019pagan} which allows the continuous and unbounded annotation of arousal using the Griffin PowerMate wheel interface. Gameplay footage was recorded at 30 frames per second (30Hz), while RankTrace provided 4 annotation samples per second (4Hz). 

\subsubsection{Maze Ball}

Maze Ball (MB) \cite{yannakakis2010towards} (or Space Maze \cite{knight2013space}) is a 3D game that served as testbed in multiple studies investigating affect detection in games \cite{yannakakis2010towards,martinez2013learning,martinez2014don,camilleri2017generalmodels,yannakakis2018artificial}. The player controls a cyan ball in a maze which contains dark ball-shaped enemies and three diamond-shaped tokens of different colors. The player has to avoid colliding with the enemies patrolling the maze, collect all the tokens and move the ball to a predefined goal point (only shown after all three tokens are collected) within 90 seconds. Each collected token adds to the player's score. The game ends either when the player runs out of time or collides with enemies twice. Background music plays during the entirety of gameplay, and sound effects play when the player obtains a token and when the player collides with an enemy.

A total of 25 players provided data for the MB dataset (the same set of players as in SS) \cite{camilleri2017generalmodels}. 
Similarly to SS, each player conducted a play-annotation cycle twice, using RankTrace and the Griffin PowerMate wheel for annotation. MB game footage was also recorded at 30Hz.

\subsubsection{Solid Rally}

Solid Rally (SR) tasks the player to drive their car through a closed circuit for two laps. In each race, the player competes against three opponents, and the goal is to finish the race on the highest possible position. Within the track, there are several checkpoints at predetermined locations: passing through a checkpoint adds to the player's score. The car engine makes sounds throughout the gameplay of Solid Rally, and a sound effect plays during car crashes. The game ends either after two laps or after 90 seconds of playing.

SR data was collected from 17 players (7 females) aged from 23 to 55 (median age 32); almost half of them (47\%) were novice or non-gamers, 35\% considered themselves expert players and the rest played games only occasionally. Each player conducted a play-annotation cycle twice using the RankTrace annotation tool provided by the PAGAN framework \cite{melhart2019pagan}. Game footage was recorded at 60Hz but downsampled to 30Hz to match the sampling rate of the other three games.

\subsubsection{Sonancia}

Sonancia (SON) \cite{lopes2019modelingsoundscapes} is a horror game taking place in a haunted dungeon divided into rooms. The players' task is to find the old statue while avoiding and outrunning monsters. The level is procedurally generated, including the number of rooms, the positioning of lights and monsters. Background audio plays throughout the game, and changes based on the room the player is in. The only sound effect is a low-volume growl when a monster sees the player.

SON data was collected from 14 players (5 females) aged from 25 to 34; 36\% of them played games everyday, 45\% played frequently or casually while the rest rarely on never played. Each participant performed a play-annotation cycle twice, using RankTrace and the Griffin PowerMate wheel for annotation. Gameplay footage duration varies from 31 to 173 seconds, and is recorded at 30Hz.

\subsection{Data cleaning}
\label{sssec:data_cleaning}

For all datasets we omit short gameplay videos with a duration under 15 seconds, in order to maintain an appropriate balance between sufficient gameplay and a player's cognitive load. This rule yields 43 videos for SS (7 short videos were omitted), 50 videos for MB, 34 videos for SR and 28 videos for SON (no video was omitted).

Since our approach is based on statistical machine learning, we explicitly assume that gameplay frames and sounds can characterize a player's arousal through an unknown mapping that machine learning aims to discover. To preserve the soundness of this assumption, we identified and omitted outlier videos whose annotations are not \emph{consistent} with the gameplay. For all games, all players play the same level, which has a specific structure, e.g. for the SS game the toys keep spawning at predetermined areas and time instances. Coupled with the fact that the duration of each session is relatively short (60, 90, 90 and maximum 173 seconds for SS, MB, SR and SON respectively), the possible states of the gameplay are restricted. Based on this observation, we assume that arousal annotation traces should, on average, exhibit a specific pattern and we can thus omit outlier videos that deviate substantially from this pattern. 

\begin{figure}[!tb]
	\begin{minipage}{0.5\linewidth}
		\centering
		\includegraphics[width=0.98\linewidth]{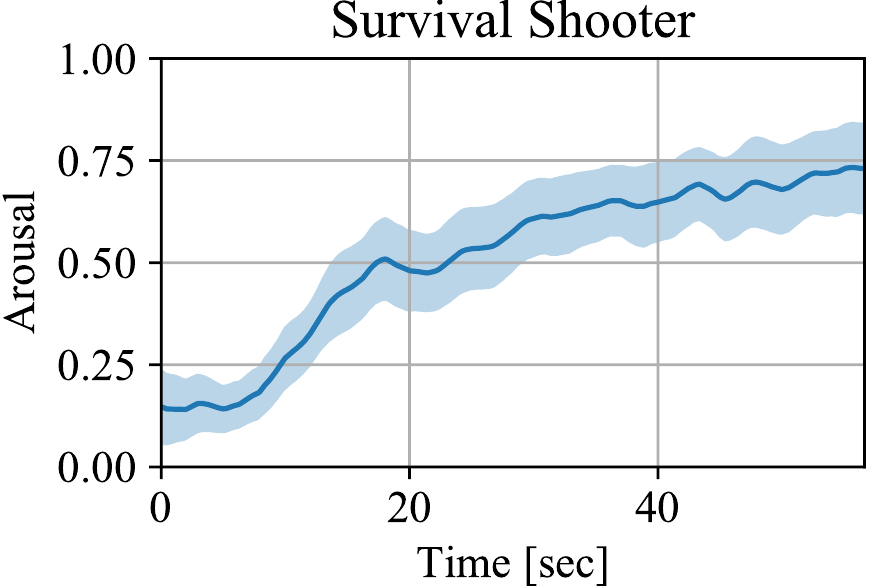}
		\vspace{0.05in}
	\end{minipage}
	\begin{minipage}{0.5\linewidth}
		\centering
		\includegraphics[width=0.98\linewidth]{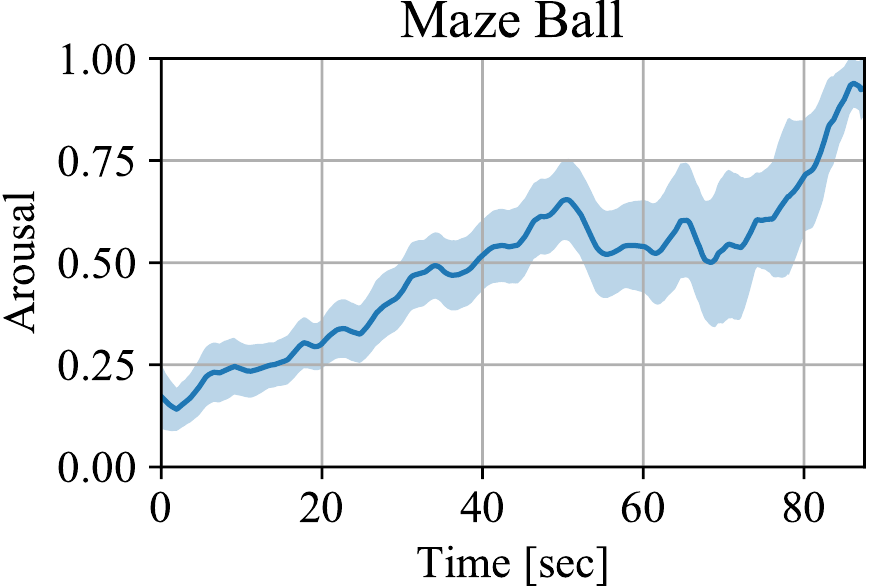}
		\vspace{0.05in}
	\end{minipage}
	
	\begin{minipage}{0.5\linewidth}
		\centering
		\includegraphics[width=0.98\linewidth]{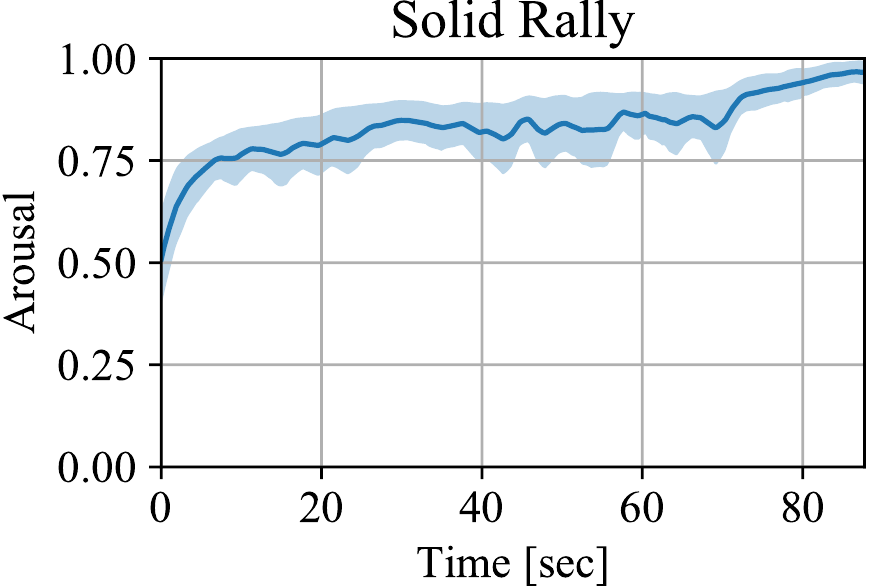}
		\vspace{0.001in}
	\end{minipage}
	\begin{minipage}{0.5\linewidth}
		\centering
		\includegraphics[width=0.98\linewidth]{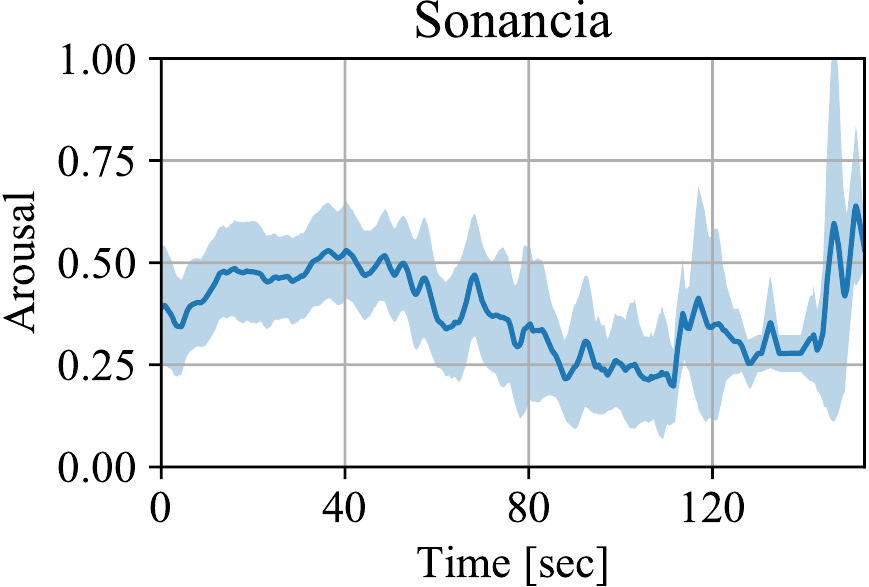}
		\vspace{0.001in}
	\end{minipage}
	
	\caption{Arousal annotation per game, averaged from signals processed at 4 Hz after min-max normalization per session. Shaded areas indicate 95\% confidence interval. As the duration of Sonancia sessions varies, the average arousal is derived from ever-fewer sessions as time progresses resulting in higher deviations of the average arousal signal.}
	\label{fig:average_traces}
\end{figure}

In particular, we denote a playthrough as an outlier if its annotation trace is dissimilar to an annotation trace that can be considered representative for the whole dataset. Since RankTrace provides continuous and unbounded arousal annotations, initially we normalise all annotation traces to $[0,1]$. Then we consider the median of all annotation traces as the representative annotation trace for the whole dataset 
and compute the distance between the annotation trace of every gameplay footage and the representative trace using the Dynamic Time Warping (DTW) \cite{berndt1994using} algorithm. DTW is widely used for measuring the similarity between two timeseries that may vary in length. The distribution of distances for each game indicates that the density is mainly concentrated on one cluster. Based on that, we omit outliers above a DTW distance threshold.
For the SS game we omit videos where the distance to the median (representative) annotation trace is larger than $0.135$; for MB, SR and SON the corresponding thresholds are $0.195$, $0.4$ and $0.2$. After removing outliers, the SS, MB, SR and SON datasets contain, respectively, 37, 45, 33 and 25 videos. Figure~\ref{fig:average_traces} depicts the average arousal trace of the cleaned dataset for each game.


\section{Learning Audiovisual Models of Affect}
\label{sec:methodology}

This study investigates the degree to which information coming from footage of the player's interaction with a game---i.e. pixels of frames and sound of a gameplay video---can act as sole predictors of a player's affective state. The RankTrace annotation tool provides continuous values of arousal, and thus it may seem natural to view the arousal estimation problem as a regression task. In this study, however, we wish to develop a user-agnostic and general approach for predicting affect without making any assumptions regarding the value of the output which may, in turn, result in biased and user-specific models \cite{yannakakis2018ordinal}. Instead, we view the challenge of arousal prediction as both a classification and a ranking task. 

This section first outlines our approach for processing the input and the output of both binary classification and preference learning models of affect (see Section \ref{ssec:training_data}) and follows with the details of the machine learning models we employ for mapping gameplay frames and sound to players' arousal (see Section \ref{ssec:model_of_affect}). 

\begin{figure}
	\begin{minipage}{1.0\linewidth}
		\centering
		\centerline{\includegraphics[width=0.97\linewidth]{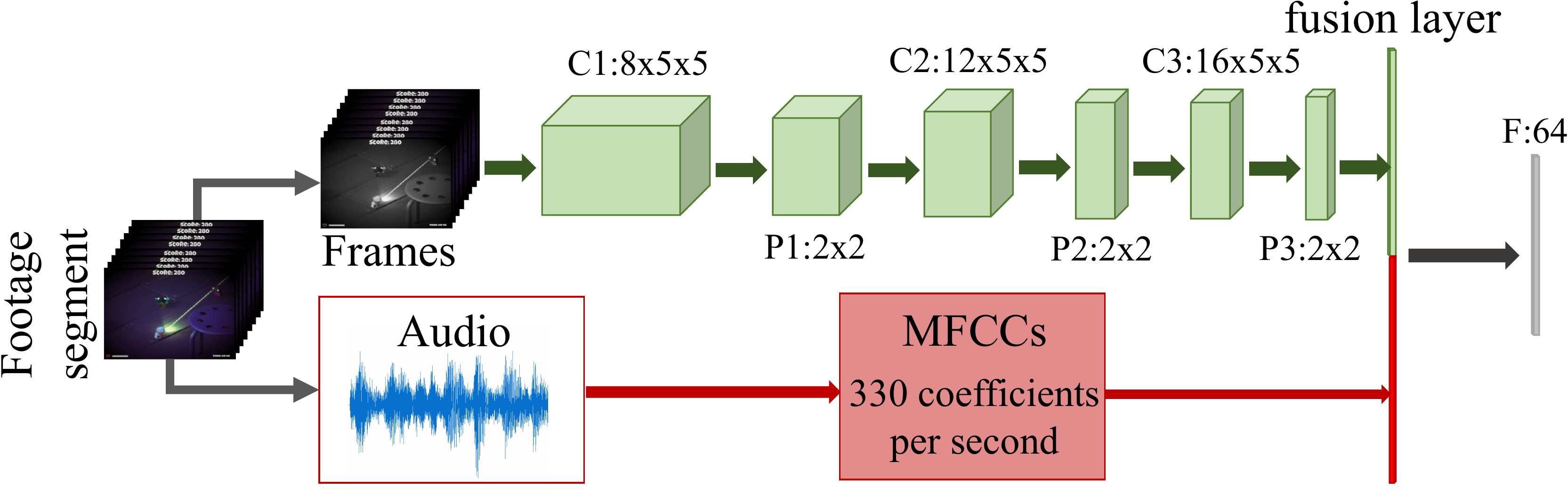}}
		\vspace{0.001in}
	\end{minipage}
	\caption{The architecture of the proposed deep learning model. The convolutional, max pooling and fully connected layers are denoted by ``C'', ``P'' and ``F'' respectively. The green stream corresponds to the processing of visual information, while the red stream to the audio information.}
	\label{fig:base_model}
\end{figure}
\subsection{Training Data Preparation} 
\label{ssec:training_data}


An obvious question that arises when a learning model is faced with \emph{video} data is how many frames it should consider. The authors in  \cite{schindler2008action} and \cite{makantasis2016deep} argue that a relatively small number of subsequent frames are adequate for representing the core elements and the content of a scene. Following this advice, we train our arousal models and evaluate their performance using small segments of footage with durations ranging from $0.25$ to $3.0$ seconds. We view the duration of a segment as a hyperparameter of both modeling approaches and we report results regarding the performance of the learning models for varying segment length. We construct those segments by splitting the videos using non-overlapping windows. The frames of those segments represent the visual information of the gameplay. To reduce the computational cost of training and evaluating the learning models, we convert the frames of gameplay videos from RGB colour to grayscale and resize them to $72\times96$ pixels for SS and MB datasets and $72 \times 115$ pixels for SR and SON datasets; doing so results in a more compact yet general-purpose representation.

As far as \emph{audio} data is concerned, we compute the Mel Frequency Cepstral Coefficients (MFCCs) \cite{molau2001computing} corresponding to the sound of each footage segment. MFCCs have been successfully used for audio classification and retrieval schemes \cite{arias2010automatic,nagavi2013content} as they can represent the spectral properties of audio data in a compact fashion. 

To construct models of \emph{arousal}, independently of the method used, we fix the range of the annotation values of each footage to $[0,1]$ using min-max normalization. Then, we synchronize the recording frequency of videos (30Hz) with annotations (4Hz) by extrapolating annotation values to each non-annotated frame.
Finally, the arousal value associated with each segment is the average of the annotation values of the frames belonging to it. 


\subsection{Deep Learning Models of Affect}
\label{ssec:model_of_affect}


To learn the unknown mapping between gameplay pixels, sounds and arousal we employ and train deep learning models to infer such a function. The deep learning models receive as input both the frames of the footage segments and the computed MFCCs and fuse those two streams of information. The learning architecture that processes and fuses the audiovisual information---for both binary classification and ranking---is depicted in Fig. \ref{fig:base_model}.


The video stream feeds a convolutional neural network that contains three convolutional layers with 8, 12, and 16 filters, respectively. The size of the filters for all convolutional layers is $5\times5$ pixels. A max-pooling operation of size $2\times2$ pixels follows each convolutional layer. The output of the convolutional stream (a feature vector of 640 elements for the SS and MB datasets and 1056 elements for the SR and SON datasets) represents the visual information of the input in a compact fashion. We should emphasize that the convolutional stream exploits both spatial and temporal information of the video frames. It exploits the spatial information by learning spatial filters (filters applied along the spatial dimension of the input). Moreover, since the learning model processes sequences of frames that exhibit temporal relations, the learned spatial filters implicitly capture and encode the temporal information of the inputs.

The audio stream receives the MFCCs as its input and propagates it directly to the fusion layer. The network does not process the MFCCs before fusing the visual and the audio streams since MFCCs are already a compact representation of the sound included in a video segment. The fusion layer, initially, concatenates the MFCCs (330 elements for each second of footage) and the features constructed by the convolutional (video) stream; it then propagates the information to a fully connected layer with 64 nodes. All aforementioned nodes use the ReLU activation function. 

All the hyperparameters of the employed model, i.e., the number and the size of hidden layers, the activation functions and the approach for fusing the two information streams, are empirically selected to balance two different criteria: (a) the computational cost of training and evaluating the model and (b) the sample complexity for avoiding under-/over-fitting. The model described above has approximately $6.5 \cdot 10^4$ trainable parameters. 

\subsubsection{Deep Classifier}\label{ssec:model_classification}

The task of arousal classification is formulated as follows; we denote $x \in \mathbb R^{h \times w \times c}$ and as $z \in \mathbb R^p$ the raw visual and audio information of a gameplay footage respectively, where $h$, $w$ and $c$ stand for height, width and length of the video segment, and $p$ for the length of the video's audio stream. Let  $\xi(x)$ and $\zeta(z)$ represent the transformations of raw information to informative features. In our case, $\xi(\cdot) \in \Xi$, where $\Xi$ denotes all possible functions that can be modeled by the CNN described in Section 4.2, and $\zeta(\cdot)$ is the function that transforms audio information to MFCCs. Having in our disposal a training set $\mathcal D =\{(x_i, z_i, y_i)\}_{i=1}^N$ of $N$ samples, where $y_i \in \{0,1\}$ for $i=1,\cdots,N$, and a class of functions $\mathcal F$ that map $\xi(x)$ and $\zeta(z)$ to $(0,1)^2$, our derived model of affect corresponds to
\begin{equation}
\label{eq:classification}
	f^*, \xi^* = \arg \min_{f \in \mathcal F, \xi \in \Xi} \frac{1}{N}L(f(\xi(x_i), \zeta(z_i)), y_i),
\end{equation}
for $i=1,\cdots,N$. $L(\cdot)$ is the negative log-likelihood loss. In our case $\mathcal F$ is the class of functions computed by feedforward fully connected networks with one hidden layer of 64 neurons and 2 output neurons activated by the softmax function (see Fig. \ref{fig:base_model}). 
The fact that we minimize the loss in (\ref{eq:classification}) with respect to both $f$ and $\xi$ indicates that our model is end-to-end trainable, i.e. the weights of the CNN (feature construction of visual input) and the classifier are optimized simultaneously.

\begin{figure}[!tb]
	\begin{minipage}{1.0\linewidth}
		\centering
		\centerline{\includegraphics[width=1.0\linewidth]{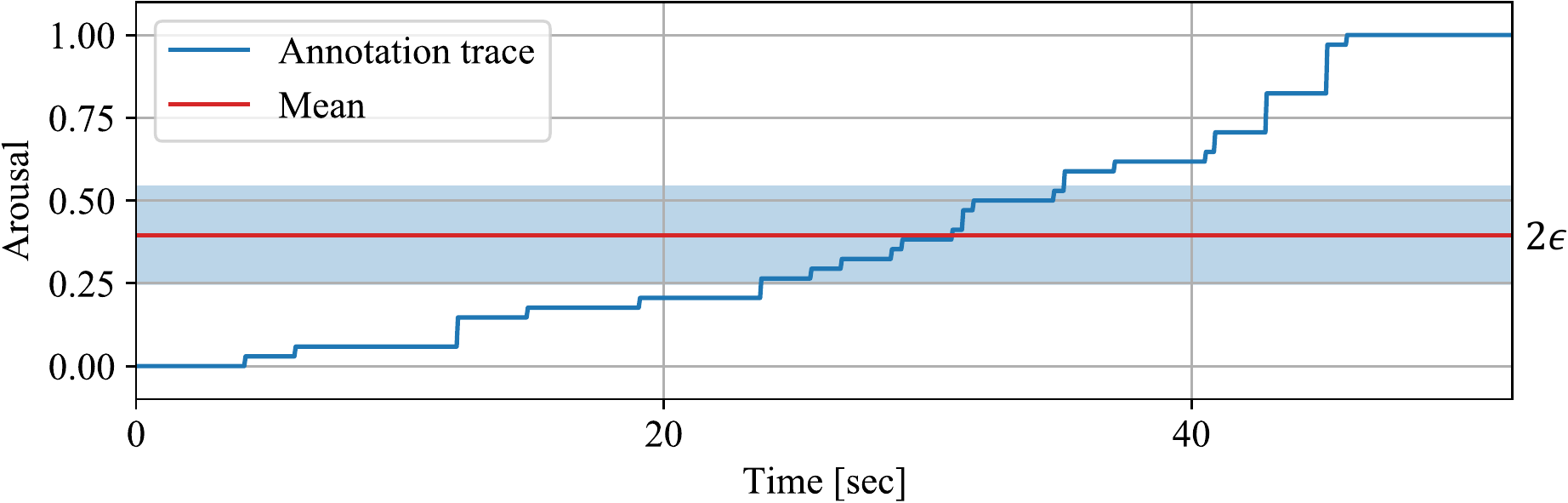}}
		\vspace{0.001in}
	\end{minipage}
	\caption{Class splitting procedure. Samples with annotations above the shaded area belong to the high arousal class, while samples with annotations below the shaded area belong to the low arousal class. The value of the uncertainty threshold $\epsilon$ defines the width of the shaded region.}
	\label{fig:class_split}
\end{figure}

For training the binary classifier  
we transform the continuous annotation values of the segments into binary values (low and high arousal) by using the mean of the annotation trace of each video as the class splitting criterion (Fig. \ref{fig:class_split}). We opt for the mean value of the annotation trace as  it is the most intuitive and unbiased way to split a continuous, unbounded annotation trace. 
Finally, we employ a threshold parameter ($\epsilon$) to determine a region around the mean within which annotation values are labeled as uncertain and ignored during classification (see the shaded area in Fig. \ref{fig:class_split}). 

\subsubsection{Deep Preference Learner}\label{ssec:model_preference}


The preference learner indicates, via its output, which one of two input segments is associated with a higher arousal value. By denoting a function $g_i(f(\xi(x_i^A), \zeta(z_i^A)) - f(\xi(x_i^B), \zeta(z_i^B)) \rightarrow (0,1)^2$ for the $i$-th $(A,B)$ pair of inputs and a dataset $\mathcal D = \{x_i^A, z_i^A, x_i^B, z_i^B, y_i\}_{i=1}^N$, of $N$ pairs, the derived preference learner corresponds to
\begin{equation}
	f^*, \xi^* = \arg \min_{f \in F, \xi \in \Xi} \frac{1}{N} L(g_i, y_i).
\end{equation}
In our case, $g(\cdot)$ is the softmax function. 
Based on this formulation, the preference learner employed here---similarly to RankNet \cite{burges2005learning,martinez2013learning}---can be seen as a binary classifier which takes as input a pair of samples and outputs 1 if the first sample in the pair is ranked higher, and 0 otherwise. 
Again, the output nodes employ the softmax activation.

Similarly to the $\epsilon$ parameter of binary classification, in preference learning we employ a threshold $\delta$ which determines if the absolute difference between the mean arousal values of two segments is high enough for the segments to be considered as a preference pair (i.e. a datapoint for training). 
Based on $\delta$ we create input pairs by comparing them in both ways; i.e. we use both $(a,b)$ and $(b,a)$ pairs, where $a$ and $b$ represent the audiovisual information of two different segments. This approach gives us a perfectly balanced dataset for deep preference learning.

\section{Experimental Results \label{sec:experiments}}

This study aims to test the hypothesis that there is a general-purpose learnable mapping of gameplay footage representation to players' affect. Towards this direction, we use the two-stream neural network (see Section \ref{ssec:model_of_affect}) for classifying game video and audio segments as high or low arousal, and for ranking them. 
For all the experiments in this paper we report the average cross-validation accuracy and the $95\%$ confidence following the demanding leave-one-video-out cross-validation scheme \cite{kearns1999algorithmic,makantasis2019pixels} which offers a highly conservative estimate for the generalization capacity of the models. 
To avoid model overfitting we employ early stopping by randomly selecting $4$ videos of the training set to form the validation set. Early stopping is activated if the classification accuracy on the validation set does not improve for $30$ training epochs. We compare the performance of the model against a baseline model which always outputs the most common class in the training set. The baseline accuracy for preference learning is always 50\%, since we have a perfectly balanced dataset (see Section \ref{ssec:model_preference}). 

\subsection{Classifying Arousal}
\label{ssec:classification}

\begin{figure}[!tb]
	\begin{minipage}{1.0\linewidth}
		\centering
		\centerline{\includegraphics[width=1.0\linewidth]{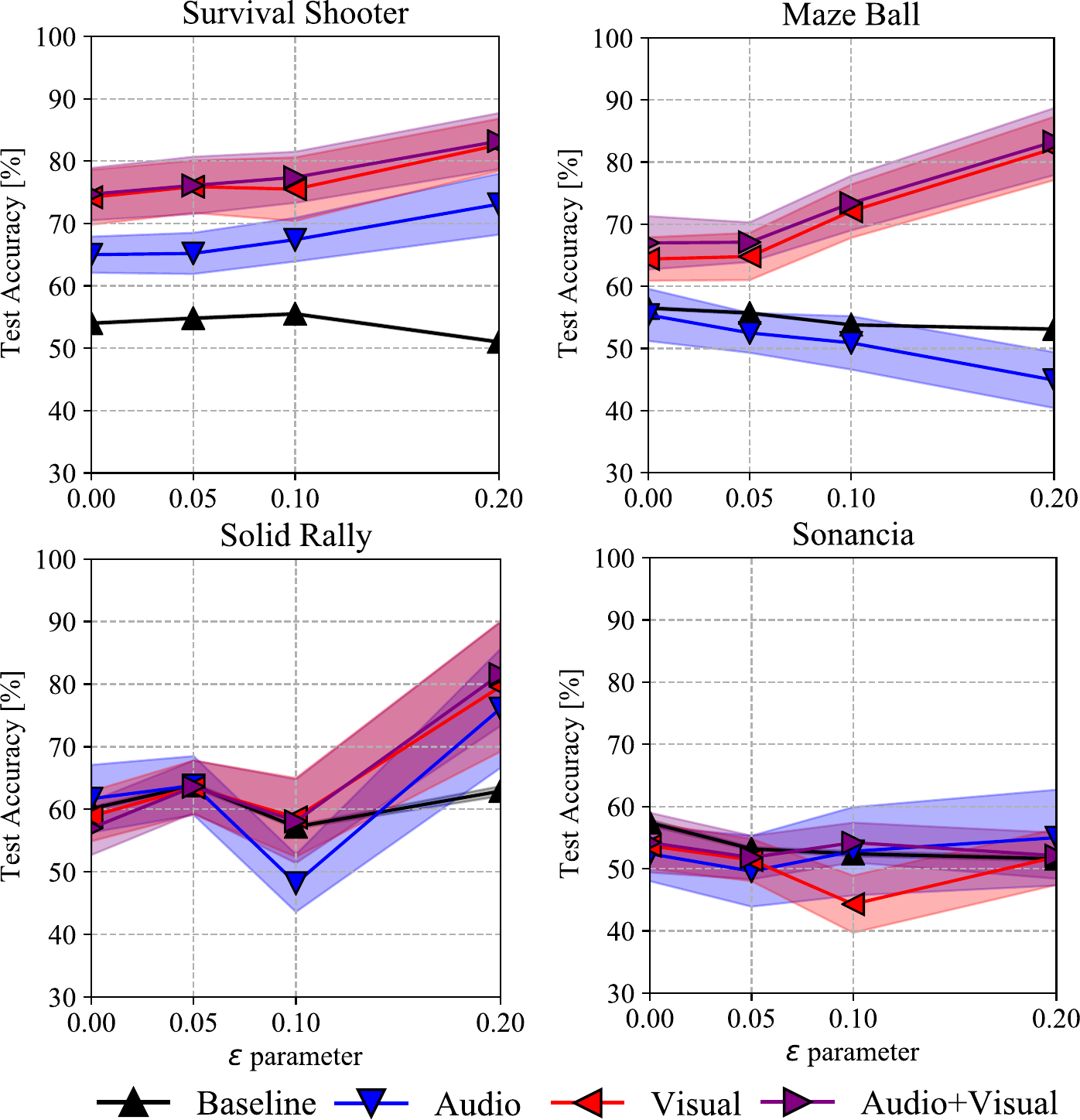}}
		\vspace{0.001in}
	\end{minipage}
	\caption{Average classification accuracy ($\%$) on the test set across the two modalities and different uncertainty threshold values ($\epsilon$). The time window is $0.5$ seconds and shaded areas indicate the $95\%$ confidence interval.}
	\label{fig:classification_tuning}
\end{figure}

To investigate the impact of the two input modalities (video frames and audio) on the performance of the model, we report the classification accuracy of three model types: two single-stream (unimodal) neural networks which are trained on either the visual or the audio information, and the two-stream bimodal neural network (see Section \ref{ssec:model_of_affect}) which is trained on both visual and audio information. Figure \ref{fig:classification_tuning} reports the average classification accuracy values obtained for different input modalities for all games, across different thresholds $\epsilon$ for omitting uncertain values near the annotation's mean value. For $\epsilon=0$, all segments of a trace are labelled high or low if their mean arousal value is above or below the mean value of the entire annotation trace ($\mu$), respectively; for $\epsilon>0$, segments with mean arousal values within [$\mu -\epsilon, \mu + \epsilon$] are omitted from the data (see Fig.~\ref{fig:class_split}).  Note that preliminary experiments established that splitting the traces into segments of 0.5 seconds (see Section \ref{ssec:window}) led to the highest accuracies across the different modalities and parameters, and as such the exploration of the best $\epsilon$ in Fig.~\ref{fig:classification_tuning} focuses on a time window of 0.5 seconds, while Table \ref{tab:dataset_size} presents the size of the employed datasets.

Unsurprisingly, the deep learning models perform better when both audio and visual inputs are considered. For SS and MB the bimodal classifier reaches accuracies as high as 30\% above the baseline classifier, but only 1\% to 3\% above the visual-only classifier. Similarly, for SR the bimodal classifier reaches accuracies 20\% above the baseline classifier and 2\% above the unimodal visual-only model. For these three datasets, most of the information regarding arousal is stored within the pixels of the video, while audio seems to play a minor role. For SS, however, audio can also be a good predictor of arousal, since even audio-only classifiers reach accuracies between 10\% and 20\% above the baseline. The same holds for SR when $\epsilon=0.2$, in which case the audio-only model reaches accuracies nearly 15\% above the baseline. On the other hand, audio-only models cannot predict arousal for the MB dataset and the SR dataset (when $\epsilon<0.2$), as accuracies are on par or worse than the baseline. For MB, audio information per se is not a reliable predictor of arousal, most likely because sound effects are rather sparse (see Section \ref{ssec:findings}). However, it can contribute to the model's predictive capacity when combined with visual information. For the SR dataset, audio information does not seem to affect the performance of the model when it is combined with visual information. For the SON dataset, both bimodal and unimodal models perform on par or worse than the baseline. In this case, neither visual nor audio information is a reliable predictor of arousal. We believe that this occurs due to the specific nature and design of the game; SON is a horror game with a delayed effect in arousal which may not be captured by the class splitting criterion (one of the limitations of our study discussed in Section \ref{sec:discussion}). As far as the design of the game is concerned, visual information in SON comes in highly vignetted frames with no HUD elements, which makes the vision-based pattern recognition task difficult and ambiguous. The background audio of the game also changes suddenly when the player moves from one room to another. Since these changes do not follow a specific pattern, audio information encoded in MFCCs cannot be easily associated with arousal.

\begin{table}[t]
	\caption{Sizes of the employed datasets for different uncertainty thresholds and time windows. For splitting the dataset into training and testing sets we follow a leave-one-video-out scheme. For the validation set (early stopping) we randomly select and use four videos in the training set.}
	\scriptsize
	\setlength\tabcolsep{1.6pt} 
	\begin{tabular}{lrrrrrrrrr}
		\multicolumn{10}{c}{\textbf{Classification}}                             \\
		& \multicolumn{4}{c|}{Time Window ($t$): 0.5 second} & \multicolumn{5}{c}{$\epsilon$=0.2}                            \\
		& $\epsilon$=0.00 & $\epsilon$=0.05 & $\epsilon$=0.1 & \multicolumn{1}{c|}{$\epsilon$=0.2} & $t$=0.25                     & $t$=0.5 & $t$=1.0 & $t$=2.0 & $t$=3.0 \\ \hline 
		SS  & $3,381$   & $3,102$   & $2,621$   & \multicolumn{1}{r|}{$1,972$}   & $3,698$                        & $1,972$   & $1,002$   & $4,83$    & $345$    \\
		MB  & $5,989$   & $5,574$   & $4,379$   & \multicolumn{1}{r|}{2852}   & $5,393$                        & $2,852$   & $1,419$   & $700$    & $448$    \\
		SR  & $4,925$   & $3,608$   & $2,225$   & \multicolumn{1}{r|}{$711$}    & $1,446$                        & $711$    & $358$    & $180$    & $119$    \\
		SON & $4,719$   & $3,707$   & $2,977$   & \multicolumn{1}{r|}{$1,846$}   & $3,474$                        & $1,846$   & $905$    & $426$    & $282$    \\ \hline \\
		\multicolumn{10}{c}{\textbf{Preference Learning}    }                                                                                   \\
		& \multicolumn{5}{c|}{Time Window ($t$): 2 seconds}                                     & \multicolumn{4}{c}{$\delta$=0.6(SS/MB), 0.4(SR), 0.75(SON)}   \\ 
		& $\delta$=0.0 & $\delta$=0.2 & $\delta$=0.4 & $\delta$=0.6                      & \multicolumn{1}{c|}{$\delta$=0.75} & $t$=0.5 & $t$=1.0 & $t$=2.0 & $t$=3.0 \\ \hline 
		SS  & $20,916$  & $13,804$  & $7,532$   & $3,860$                        & \multicolumn{1}{r|}{$2,138$}   & $67,576$  & $16,380$  & $3,860$   & $1,584$   \\
		MB  & $43,090$  & $25,740$  & $13,854$  & $6,072$                        & \multicolumn{1}{r|}{$2,766$}   & $104,258$ & $25,330$  & $6,072$   & $2,488$   \\
		SR  & $39,813$  & $14,146$  & $4,898$   & $2,324$                        & \multicolumn{1}{r|}{$1,358$}   & $41,588$  & $9,940$   & $4,898$   & $888$    \\
		SON & $43,844$  & $25,108$  & $10,364$  & $3,502$                        & \multicolumn{1}{r|}{$1,282$}   & $26,527$  & $6,509$   & $1,282$   & $532$    \\ \hline
	\end{tabular}
	\label{tab:dataset_size}
\end{table}


In summary, for 3 out of 4 datasets, the high performance obtained by varying the uncertainty bound indicates that the mapping between general-purpose representations of audiovisual gameplay information and arousal can be learned statistically with very high accuracy. Results for the SON dataset indicate that the performance of our models depends on the specific nature of the game, as well as on the underlying assumptions of our approach (see Section \ref{sec:discussion}).


\subsection{Ranking Arousal}
\label{ssec:ranking}

\begin{figure}[!tb]
	\begin{minipage}{1.0\linewidth}
		\centering
		\centerline{\includegraphics[width=1.0\linewidth]{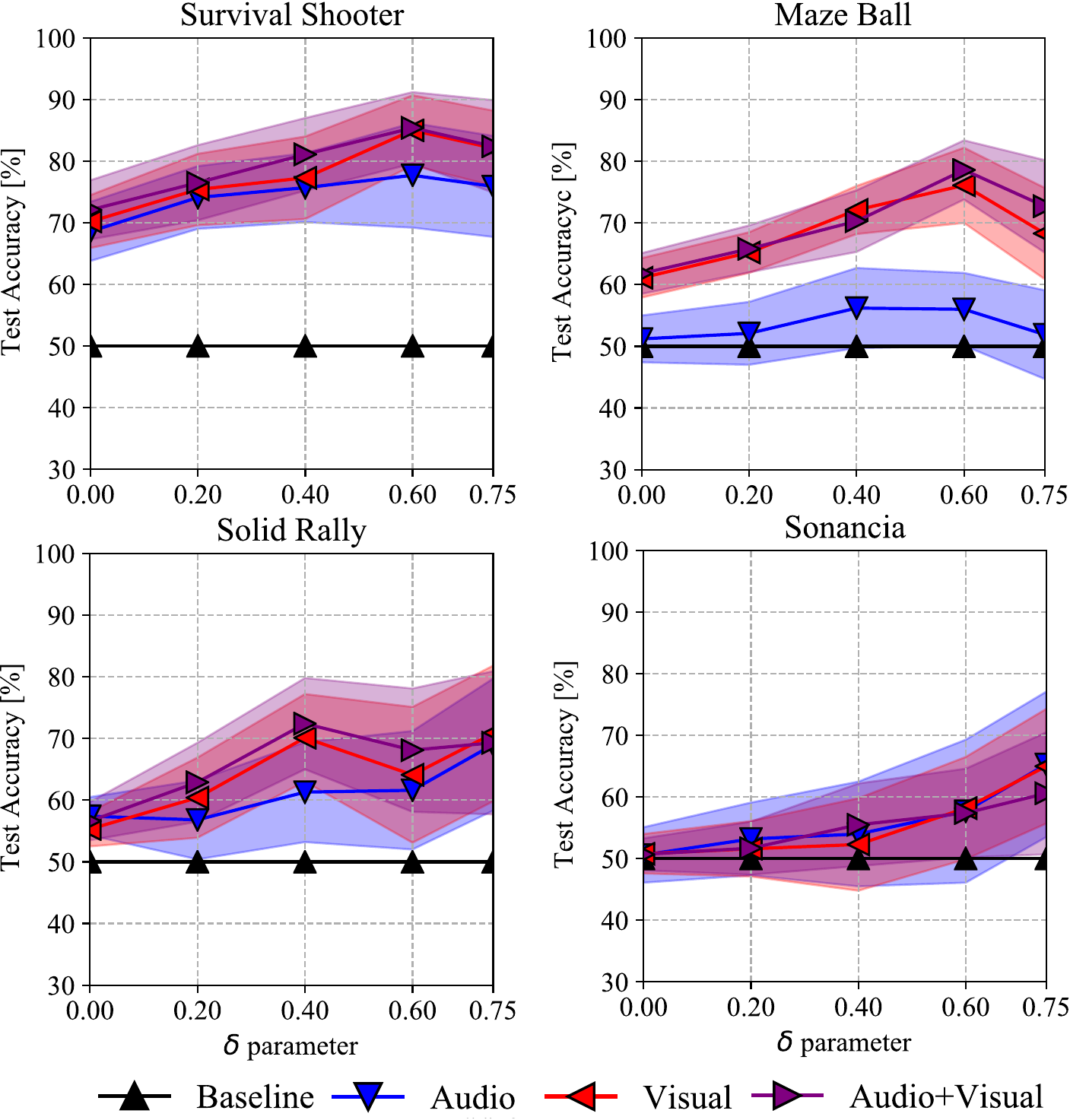}}
		\vspace{0.001in}
	\end{minipage}
	\caption{Average accuracy ($\%$) of the preference learner on test set across the two modalities and different uncertainty threshold values ($\delta$). The time window is $2.0$ seconds and shaded areas indicate the $95\%$ confidence intervals.}
	\label{fig:preference_tuning}
\end{figure}

Similar to Section \ref{ssec:classification}, we investigate how different modalities affect the performance of the preference learner by training three preference learning models with different inputs. The preference learner compares two input segments and outputs which segment has a higher arousal value. Based on preliminary experiments we focus on the best time window for the preference learner, which is 2 seconds.


The $\delta$ parameter (as defined in Section \ref{sec:methodology}) sets the minimum absolute difference between the mean annotation value of two segments that can be considered as a preference. In this section, we investigate the performance of the preference learner in terms of average classification accuracy for 5 different values of $\delta$, i.e., $\delta=\{0.0, 0.2, 0.4, 0.6, 0.75\}$, and for the different input modalities. For all the experiments presented, we follow the leave-one-video-out validation procedure using segments of 2 seconds. Figure \ref{fig:preference_tuning} summarises the results of this investigation, and the sizes of the datasets are presented in Table \ref{tab:dataset_size}. 

The preference learner achieves up to 32\%, 28\%, 22\% and 11\% higher accuracy than the random baseline for the SS, MB, SR and SON dataset, respectively. As with the classifier, the models that exploit both audio and visual input perform better than unimodal models. While high values for $\delta$ yield pairs of inputs that have significantly different annotation values, this also results in smaller datasets. According to Fig.~\ref{fig:preference_tuning}, for SS and MB we obtain the highest performance values when $\delta= 0.6$, for SR when $\delta=0.4$ and for SON when $\delta=0.75$. These threshold values seem to balance between highly informative and comparable inputs, and adequately large dataset size for training (see Table \ref{tab:dataset_size}). 

\subsection{Impact of the Time Window}
\label{ssec:window}

In all experiments presented so far we investigated the performance of the arousal model by keeping the time window of the input signal constant ($0.5$ seconds for classification and $2$ seconds for preference learning). In this section, we vary the time window while retaining the best $\delta$ and $\epsilon$ values found in Sections \ref{ssec:classification} and \ref{ssec:ranking} respectively. We assume that the duration of the gameplay videos affects the model performance for three reasons: first, the length of the window determines directly the size of the dataset; second, the duration of footage segments determines the amount and the quality of the audiovisual information contained in a segment (i.e. the longer the segment, the richer the information contained in it); third, the duration of the window affects the ground truth arousal values as those are averaged from the window's annotation trace.  

\begin{figure}[!tb]
	\begin{minipage}{1.0\linewidth}
		\centering
		\centerline{\includegraphics[width=1.0\linewidth]{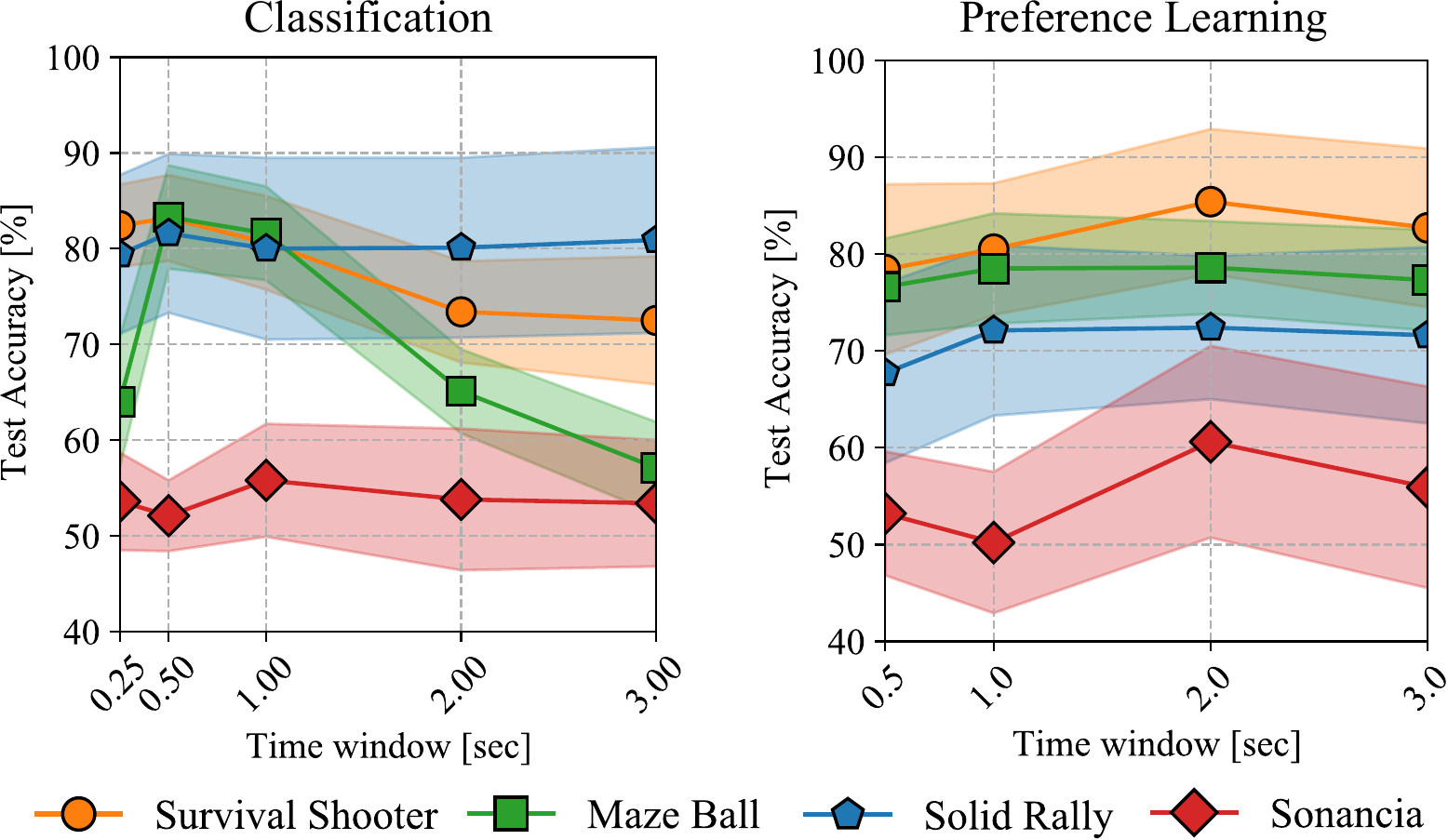}}
		\vspace{0.001in}
	\end{minipage}
	\caption{Average accuracy ($\%$) on test set for the best audiovisual model across different time windows, for classification and preference learning. Shaded areas indicate the $95\%$ confidence intervals.}
	\label{fig:time_sensitivity}
\end{figure}

Figure \ref{fig:time_sensitivity} (left) summarizes the impact of window duration on the accuracy of our proposed two-stream (audio and visual) neural network for the classification task. For all results, the uncertainty threshold value is fixed to its best value: $\epsilon=0.2$. For the SS game, accuracy is consistently over $70\%$ for all durations. However, the model achieves the best performance for $0.5$ second windows, and accuracies drop for longer windows. It appears that the fast pace of the game does not favor inputs of long duration since their ground truth annotation values are over-smoothed. For the MB game, the accuracies deviate wildly in different time windows. In particular, the performance of the model is over $80\%$ for segments of 0.5 or 1 seconds, and less than $65\%$ for shorter segments ($\sim$0.25 second). Contrary to SS, MB is a slow paced game. Therefore, in this game it seems that short segments do not contain sufficiently rich information for the classification task, and thus do not contribute towards the efficient training of the model. Both games (especially MB) perform worse in segments over 1 second, also due to the fact that the size of the dataset becomes too small for training (for 3 seconds and $\epsilon=0.2$ the datasets for SM and MB are only 345 and 448 segments). For the SR dataset the classification accuracy is $\sim$80\% for all time windows. However, it shows wide confidence intervals due to the small number of training samples (as shown in Table \ref{tab:dataset_size}, for 2 or 3 second windows less than 200 samples are retained). 

Figure \ref{fig:time_sensitivity} (right) similarly visualizes the impact of time window length on the accuracy of the preference learner using audiovisual input, and with the best $\delta$ value ($\delta=0.6$ for SS and MB, $\delta=0.4$ for SR and $\delta=0.75$ for SON). As indicated in Section \ref{ssec:ranking}, the best performance for both datasets is obtained for segments of 2 seconds. For the SS dataset the learner that uses 2 second segments as input performs $5\%$ and $2.7\%$ better that the models that use 1 and 3 second segments, respectively. For the MB and the SR games, all models perform almost the same irrespective of the time window considered, although for short time windows (0.5 seconds) the performance drops in both games. For SON the preference learner performs best for 2 seconds time windows ($11\%$ above the baseline), and 3 seconds ($6\%$ above the baseline). This suggests that Sonancia, as a horror game, elicits affect in a delayed fashion and thus requires longer segments of gameplay to be considered. Due to the way that preference learning processes the dataset, the number of preferences increases exponentially compared to the number of segments themselves: based on Table \ref{tab:dataset_size}, the number of preferences at 0.5 seconds are in the power of $10^4$ (up to 50 times the number of preferences at 2 seconds). 
For segments of 0.25 seconds, the dataset explodes and training becomes problematic due to computational effort. 

\subsection{Classification vs. Ranking}
\label{ssec:versus}

In this section we compare the preference learner against the binary classifier. While both methods yield high accuracies for three of the four games, such a metric is not appropriate for conducting a fair comparison between the methods as the set of input-output pairs for the two approaches is not the same \cite{martinez2014don}. 
Following the method introduced by Martinez \emph{et al.} \cite{martinez2014don}, we compare the two approaches based on the global orders of arousal they produce when they are fed with inputs that belong to the same gameplay footage. The orders produced by the models are evaluated against the ground truth global order which is derived by the arousal annotation values. Inspired by \cite{martinez2014don, yannakakis2018ordinal} we compare the methods using the Kendall's rank correlation coefficient ($\tau$), which measures the ordinal association between two rankings \cite{abdi2007kendall}. We calculate $\tau$ on the test video in a leave-one-video-out cross-validation process, and report the 95\% confidence intervals across all videos in each dataset.

\begin{figure}
    \centering
    \includegraphics[width=0.47\textwidth]{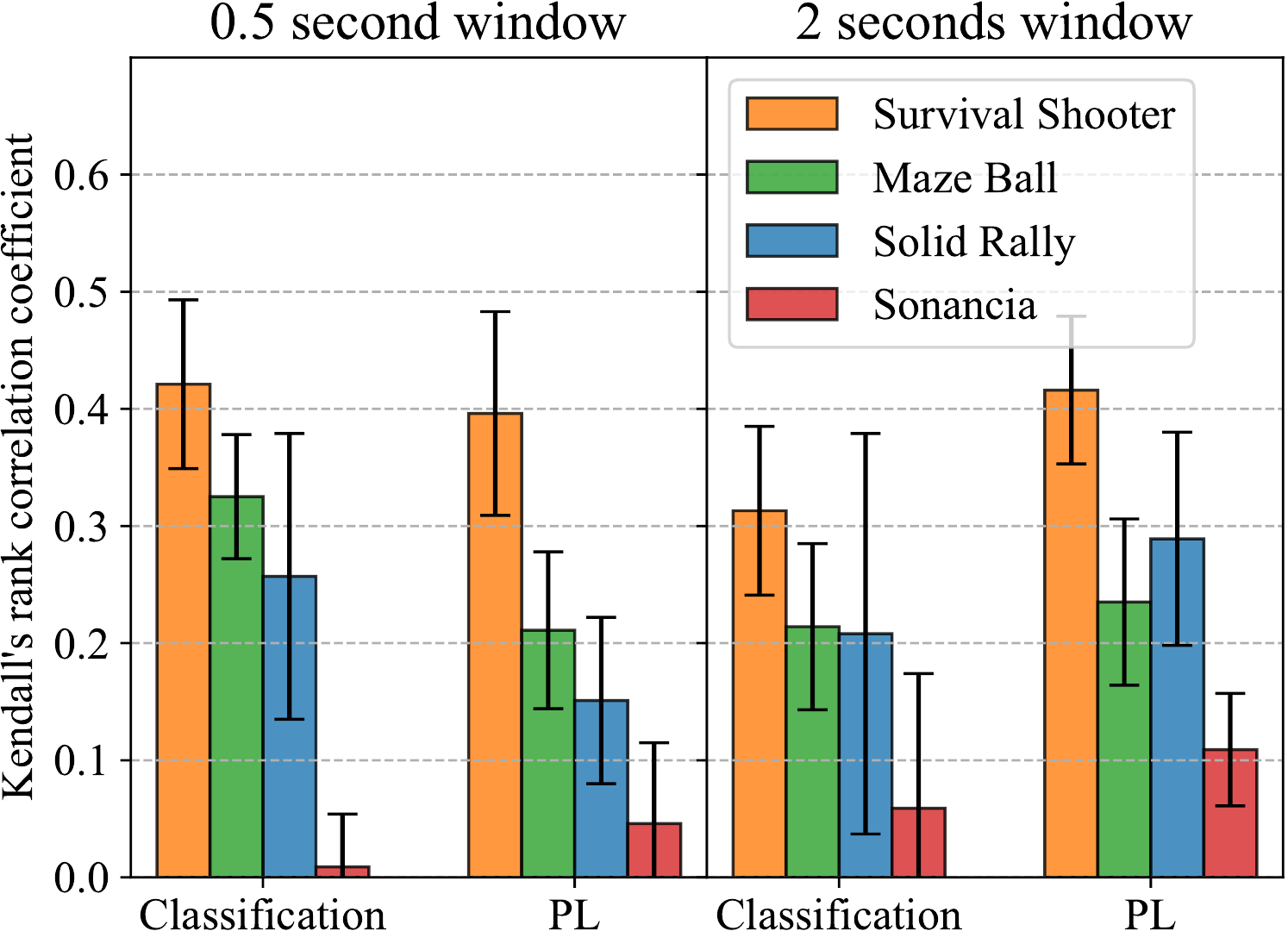}
    \caption{Kendall's $\tau$ across modelling approaches, games, and best time windows for each approach. Values are averaged from the leave-one-video out cross-validation and error bars denote the 95\% confidence interval.}
    \label{fig:tau}
\end{figure}

For both approaches we use the trained models presented in Sections \ref{ssec:classification} and \ref{ssec:ranking} which achieve the best classification accuracy: for classification, the best models are with $\epsilon=0.2$ and a time window of $0.5$ seconds for all games, and for preference learning the best models are with $\delta=0.6$ for SS and MB, $\delta=0.4$ for SR and $\delta=0.75$ for SON and a time window of 2 seconds. Fig.~\ref{fig:tau} also shows the $\tau$ values for models trained on both time windows for both approaches, for a more holistic comparison. The average Kendall's $\tau$ for both datasets indicates---unsurprisingly---that the produced orderings are positively correlated to the ground truth independently of the method used. For the SS and SR games both approaches seem to perform almost the same for their best models (with no significant differences). For MB the classifier yields higher $\tau$ values than the preference learner and for SON the best preference learner yields higher $\tau$ values than the best classifier (which is also at 2 seconds), but these differences are not statistically significant. 
As evident from Section \ref{ssec:window}, the classifier performs worse at 2 seconds windows (except for SON) while preference learning performs worse at 0.5 second windows compared to each method's optimal time window. It should be noted that the way classification and preference learning process the data results in a very different treatment (classes versus ranks) which makes a completely fair comparison very difficult. Indicatively, classification with 2 second windows and $\epsilon=0.2$ operates on a dataset of size $483$, $700$, $180$ and $426$ for SS, MB, SR and SON respectively, versus $3,860$, $6,072$, $4,898$ and $1,282$ for preference learning (with the best $\delta$ values per game). Therefore, using the best models for each approach even if the input is different (specifically, the number of frames used as input to the CNN, and the number of MFCCs for audio) is the most straightforward comparison as the number of samples (with the chosen $\epsilon$, $\delta$ parameters) are in the same order of magnitude.

Based on the comparison above, we conclude that a binary classifier can reach comparable accuracies to the preference learner, or higher in the case of MB. The accuracy of the binary classifier comes at the cost of the resolution of the output (as there are only two classes). If the problem requires larger output resolution (e.g., high, medium and low arousal), it is not clear how a 3-class classifier could produce such orderings. On the other hand, preference learning models can always produce orderings via pairwise comparisons of inputs and they appear to be more robust across time windows and across all games tested.

\subsection{Analysis of Findings}
\label{ssec:findings}

The experiments presented in this paper showed that it is possible to construct accurate models of players' arousal based on general-purpose representations of gameplay footage. The results obtained across different input modalities also indicate that the visual information is key for the efficiency of the models. Moving towards higher degrees of model expressivity and explainability, in this section we attempt to identify the features of the gameplay video that contribute more to the output of the arousal models. One way to achieve this is by visualizing the areas of the frame that have the highest influence on the model's prediction. For that purpose we use the Gradient-weighted Class Activation Mapping (GCAM) method \cite{selvaraju2017grad}. For any given input, GCAM computes the gradient of the output neuron with respect to the neurons of a convolutional layer. By multiplying the given input with the computed gradient, we obtain a heatmap that indicates how much each area of the input contributes to the output.

\begin{figure}[!tb]
\centering
	\begin{minipage}{1.0\linewidth}
		\centering
		\centerline{\includegraphics[
		width=1.0\linewidth]{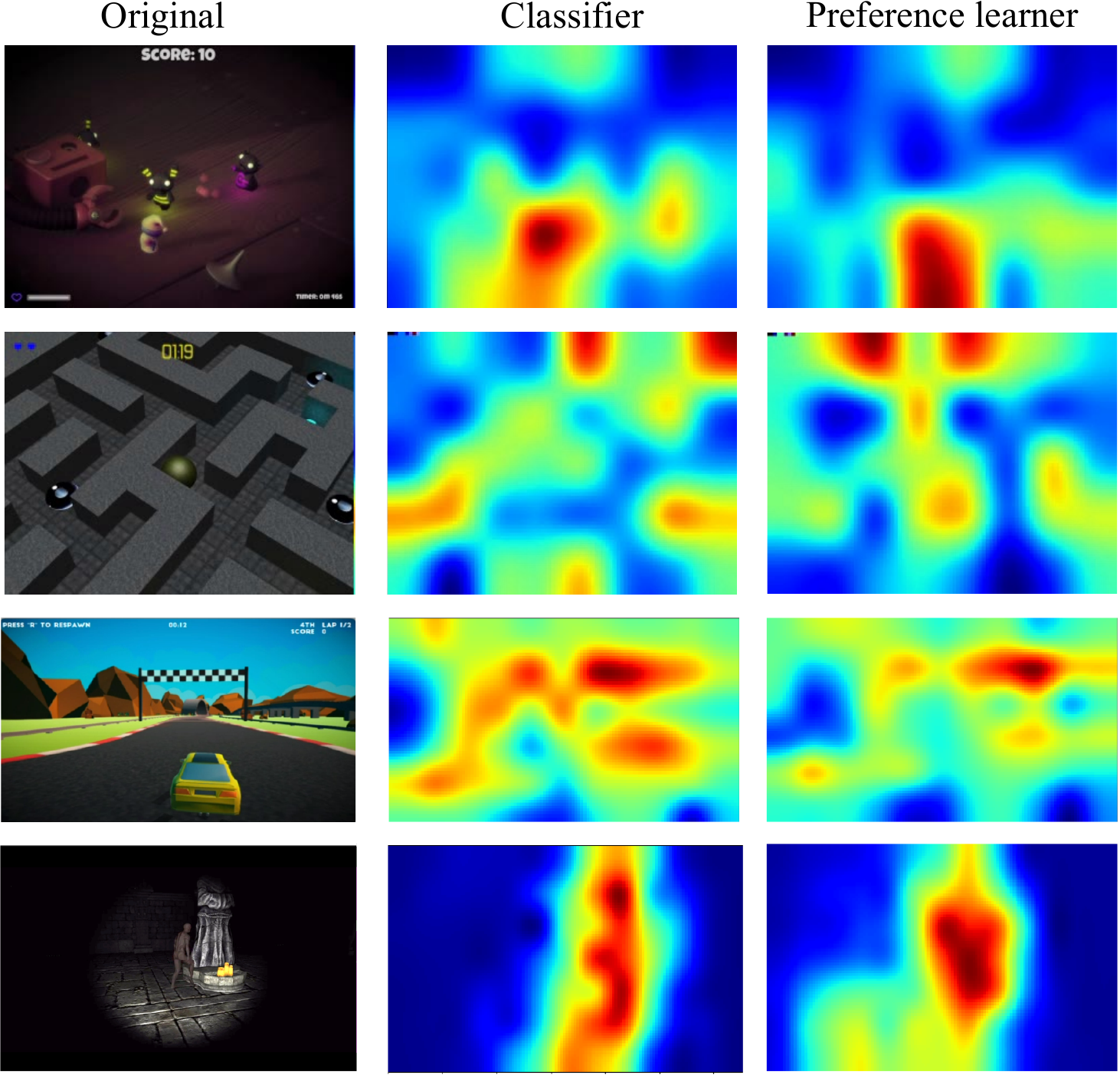}}
		\vspace{0.001in}
	\end{minipage}
	\caption{Activation maps of a sample footage segment for all games. We display the last frame of the segment.}
	\label{fig:activation}
\end{figure}

Figure \ref{fig:activation} depicts the activation maps for a sample footage segment for different games and learning paradigms; for visualisation purposes, we display the last frame of the segment. We observe that aspects of the heads-up display (HUD) affect the arousal prediction. For SS, the score located at the top centre of the screen---which keeps increasing during the progression of the game as the player kills more and more hostile toys---contributes significantly to arousal prediction. The pixels of the avatar and the hostile toys, however, seem to have the highest impact on the outcome independently of the method used. Similarly, the time indicator on the HUD of the MB game contributes highly to the arousal prediction regardless of the learning paradigm used. Besides the time indicator, the location of the ball, the enemies and the tokens appear to have a substantial impact on arousal prediction. Interestingly, the HUD element of the player's health was not considered for either game. For SR, the HUD elements and the player's avatar (car) do not seem to be important features for either approach; the focus is instead on level elements immediately in front of the car such as the finish line or the loop in the horizon. For SON, it is obvious that the lack of HUD and dark visuals (only a small part of the screen contains information) confuse the classifier, although both approaches identify the statue (the goal of the game) and the monster as important features.

As a general comment from our qualitative analysis, there are two key differences in how sound influences the arousal models in different games. On the one hand, sound effects in SS follow shooting and enemy deaths which are common events and information-rich (e.g. killing an enemy means that the player survives longer), while for MB sound effects are rare since they trigger when a token is picked (with three tokens in the game) or when the player dies (which will not be a common event). Sound effects are more common and can thus be exploited better in SS, which explains the low performance of audio channels in arousal detection for MB. SR on the other hand has a persistent sound from the engine, making some frequencies in the MFCC near-constant. Finally, in Sonancia there are almost no sound effects (only if a monster sees the player) and the background audio changes based on the room in a non-diegetic way \cite{ekman2005noise}. The sound design in SON is thus expected to confuse the models of affect as the soundscape variation is sparse and only captures the gameplay context very indirectly.

\section{Discussion}
\label{sec:discussion}

The most common approaches for modelling affect rely on direct observations and measurements of human behaviour. The proposed study, to the best of our knowledge, presents one of the first attempts to model affect via general-purpose representations of information that comes solely from gameplay footage that does not display human behaviour directly. Human behaviour is embedded into the gameplay footage, e.g. as avatar movement and actions, since emotion is manifested through and annotated on the video per se.

We exploit bimodal audiovisual information to build and test a deep neural network for predicting players' arousal states via two different approaches; binary classification and preference learning. Results show, on the one hand, that building such models is possible, and on the other, that these models can be highly accurate in most cases. Moreover, more robust and accurate models can be constructed when the dataset is pruned from ambiguous data. 

In this study, we make several assumptions which implicitly indicate possible limitations of our approach. First, we use the mean value of each annotation trace to split data into high and low arousal classes. Although this criterion is intuitive and straightforward, it makes sense only for stationary processes. In our study, this criterion results in robust annotations due to the short duration of gameplay. For long playthroughs, however, this assumption will not hold and using the mean as the class splitting criterion may produce misleading classification and preference learning results. Second, we use a representative annotation trace (median trace) to detect and remove outliers. In other words, our data cleaning methodology considers only the distribution of the annotation values. In our study, such a methodology can efficiently detect and remove outliers since the games considered can be played in specific ways, and gameplay duration is short. For sandbox games or long play sessions, a data cleaning methodology that takes into consideration simultaneously the input and the output distributions should be used (i.e., the joint distribution of audiovisual information and annotation values). Finally, the data points used for training the affect models are generated sequentially. Thus the annotation of a data point at a specific time instance might depend on the annotation value of the data point generated before. Our models, however, are not able to exploit this information. To take advantage of this kind of information, models that explicitly take into consideration the temporal ordering of data, such as LSTMs, should be used.

The differences in performance among the four games also illuminate some concerns regarding the impact of the game environment on the feasibility of general-purpose models. As discussed in Section \ref{sec:dataset} and Section \ref{ssec:findings}, each game is different in terms of what the player sees (camera perspective, color scheme, illumination), hears (background audio, variety and volume of sound effects) and performs (control schemes, actions per minute, degree of immediate feedback, available actions, clarity of game goal). Based on these differences, it is expected that the player also feels (and annotates) differently in each game (see Figure \ref{fig:average_traces}). While in games such as Survival Shooter highly accurate models of affect could be trained via either approach, in Sonancia specifically the performance of the classifier was not better than the baseline and the preference learner could reach accuracies of $\sim60\%$ with the best parameter setup. It can be gleaned that arcade games with fast-paced interactions (such as SS and SR), a top-down camera perspective that shows more of the level (SS and MB), distinct forms and colors to distinguish game objects (MB and SR), loud sound effects tied to game events (SS) could help the model predict affect from the audiovisual signals alone. In contrast, Sonancia has none of the above design patterns; moreover, the actions that a player takes (e.g. choosing a room to go into) do not have immediate gameplay (and, one would assume, affective) impact as a monster could be hiding in a remote part of the room she chooses to go. Future work should explore where the limits are in terms of game environments and visual, audio or interaction design for which this method can be applied. While it is expected that high-contrast and fast-paced arcade games such as the ATARI games studied by Mnih \emph{et al.} \cite{mnih2015human} will work very well for this method, it is unclear whether audiovisual signals in time windows of a few seconds would work well for e.g. role-playing games (which require long interactions), visual novels (where the story consequences are not displayed visually or immediately) or turn-based games (where real-time windows are irrelevant). Exploring how these different design patterns affect the quality of predictive affect models based on audiovisual data alone can be useful not only for affective computing but also for game design, as it can inform designers how to maximize the emotional impact of their content.

While this study is one of the first attempts at the challenging task of predicting affect states from general-purpose gameplay footage information, the results are promising and point to a number of extensions in future work. 
In this paper, our models require training on each particular game; while the method is robust and general-propose, the models themselves have not been tested for their generality. To test for the model's generality, a future direction would be to devise leave-one-game-out validation schemes once our game corpus becomes even larger. Such a cross-validation scheme would allow us to test the degree to which certain characteristics of audiovisual information are general predictors of arousal and transferable to other games. In terms of the model's input, we use grayscale frames to represent the visual information and MFCCs for the audio information. Grayscale frames and MFCCs can compactly represent the audiovisual information of gameplay footage and reduce the computational cost of training the models. These representations, however, can be enhanced without sacrificing the generality of our approach. In terms of sounds, MFCCs can be fused with the concise GeMAPS feature set \cite{eyben2015geneva} which has been successfully used for voice recognition and affective computing applications. As far as the representation of visual information is concerned, it can be enhanced by using RGB channels or hand-crafted channels that include low-level image information \cite{makantasis2015deep}. For example, exploiting hue and saturation information could better detect the red monsters present in Sonancia. In terms of output (affect labels), we use the mean arousal value within a time window. That is an intuitive approach, which, however, can be further investigated and refined. For example, amplitude and average gradient \cite{lopes2017ranktrace, camilleri2017generalmodels} could be used for processing annotations within a time window. 

Beyond arousal, the method's robustness needs to be tested for other affective dimensions---including valence and dominance---or continuous affective states such as engagement \cite{melhart2020moment}. Beyond games, the method appears to be generalizable to any rich human computer interaction domain that interweaves the context of interaction with user behavior and user affect, such as mobile app interaction and web navigation. Additional experiments in datasets of that type, however, need to be performed to validate this hypothesis.   

\section{Conclusions}
\label{sec:conclusions}

In this paper we introduced a general methodology for predicting affect solely from audiovisual aspects of human computer interaction. Our hypothesis is that arousal embedded in affective interaction can be modeled accurately without considering any user manifestation of affect besides the pixels and the sound of the interaction. 
The hypothesis was tested in digital games, a domain that interweaves affect with audiovisual content through gameplay interaction. The audiovisual content in games has a dual role: it is both the elicitor of affect and the context of the interaction. We developed two deep learning paradigms for mapping directly from pixels and audio of videos to the annotated arousal of gameplay: a deep classifier and a deep preference learner, both using a combination of CNN and feedforward architectures. Our experimental results across four dissimilar games suggest that arousal can be predicted with very high accuracies via such general-purpose representations (as high as 85\%) as long as the audiovisual feed captures the gameplay context accurately (which depends on the game's design). The fusion of the two modalities (gameplay pixels and sounds) unsurprisingly appears to be beneficial for the predictive capacity of the models. Our key findings also show that activation maps can visualize the areas on the screen that lead to high arousal---in our case primarily the score, the avatar and the enemies. The GCAM visualization increases the explainability of the models \cite{zhu2018explainable} and can be very useful for game designers when adjusting the appearance or in-game function of the game elements to increase or decrease the elicited emotion of certain events.

This paper defines one of the first steps towards the creation of general representations of affect by studying arousal detection in games. The results showcase that it is possible to detect arousal accurately by only considering low-level contextual information of the interaction. The key findings are relevant to any application area within affective computing and directly applicable to domains of rich human computer interaction that consider user affect.

\ifCLASSOPTIONcompsoc
  \section*{Acknowledgments}
\else
  \section*{Acknowledgment}
\fi
We would like to thank the anonymous reviewers for their constructive feedback and detailed comments that helped us improve the quality of the paper. Konstantinos Makantasis was supported by the European Union's H2020 research and innovation programme 
(Grant Agreement No. 101003397). Antonios Liapis and Georgios N. Yannakakis were supported by the European Union's H2020 research and innovation programme 
(Grant Agreement No. 951911). 

\ifCLASSOPTIONcaptionsoff
  \newpage
\fi

\begin{IEEEbiography}[{\includegraphics[width=1in,height=1.25in,clip,keepaspectratio]{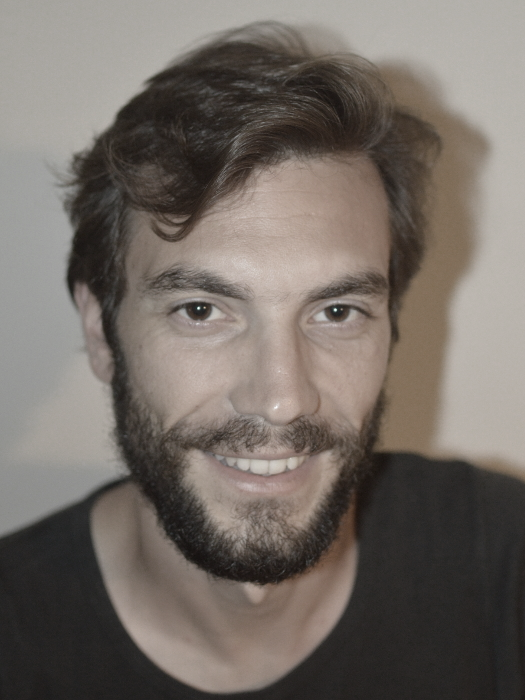}}]{Konstantinos Makantasis} received his computer engineering diploma and his Master degree from the Technical university of Crete (TUC, Greece). In 2016 Dr. Makantasis received his PhD from the same school working on detection and semantic analysis of objects and events through visual cues. Currently, he is a MSCA IF Widening Fellow at the Institute of Digital Games, University of Malta working on tensor-based machine learning methods for affect modeling. He is mostly involved and interested in computer vision, machine learning / pattern recognition and probabilistic programming. He has more than 45 publications in international journals and conferences on computer vision, signal and image processing and machine learning.
\end{IEEEbiography}

\begin{IEEEbiography}[{\includegraphics[width=1in,height=1.25in,clip,keepaspectratio]{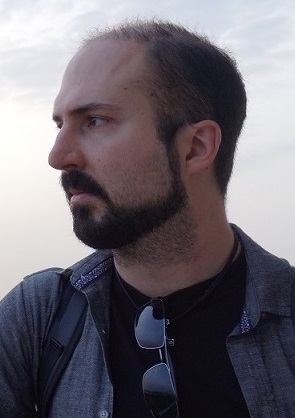}}]{Antonios Liapis} is a Lecturer at the Institute of Digital Games, University of Malta, where he bridges the gap between game technology and game design in courses focusing on human-computer creativity, digital prototyping and game development. He received the Ph.D. degree in Information Technology from the IT University of Copenhagen in 2014. His research focuses on Artificial Intelligence as an autonomous creator or as a facilitator of human creativity. His work includes computationally intelligent tools for game design, user models for the design process, gameplay, or visual preference, as well as evolutionary computation and deep learning. He has published over 100 journal, conference and workshop papers on these topics, and has received several awards for his research contributions and reviewing effort. He has served as general chair in four conferences (EvoMusArt 2018-2019, GALA 2019, FDG 2020), and as a Guest Editor in three special issues in {\sc IEEE Transactions of Games}, ACM Journal on Computing and Cultural Heritage, and the International Computer Games Association Journal. 
\end{IEEEbiography}


\begin{IEEEbiography}[{\includegraphics[width=1in,height=1.25in,clip,keepaspectratio]{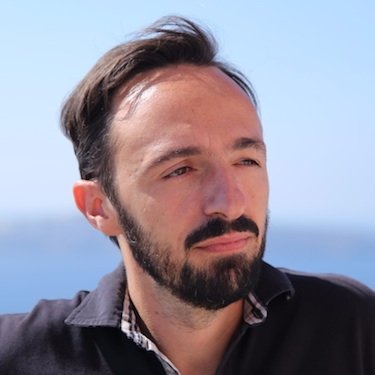}}]{Georgios N. Yannakakis}(S'04-M'05-SM'14) is a Professor and Director of the Institute of Digital Games, University of Malta, and a co-founder of modl.ai. He received the Ph.D. degree in Informatics from the University of Edinburgh in 2006. He does research at the crossroads of artificial intelligence, computational creativity, affective computing, advanced game technology, and human-computer interaction. He has published more than 260 papers in the aforementioned fields and his work has been cited broadly. He is currently an Associate Editor of the {\sc IEEE Transactions on Games} and the {\sc IEEE Transactions on Evolutionary Computation} and used to be Associate Editor of the {\sc IEEE Transactions on Affective Computing} journal. He has been the General Chair of key conferences in the area of game artificial intelligence (IEEE CIG 2010) and games research (FDG'13, '20). Among the several rewards he has received for journal and conference publications he is the recipient of the \emph{IEEE Transactions on Affective Computing Most Influential Paper Award} and the \emph{ACII 2017 Best Paper Award}. He is a senior member of the IEEE.
\end{IEEEbiography}




\end{document}